\documentclass[journal=jacsat,manuscript=article]{achemso}
\setkeys{acs}{articletitle=true}

\usepackage{chemformula} 
\usepackage[T1]{fontenc} 
\usepackage{multirow}
\usepackage{bigdelim}
\usepackage{longtable}
\usepackage{graphicx}
\usepackage{times}
\usepackage{amssymb}
\usepackage{achemso}

\newcommand\arcsec{\hbox{$^{\prime\prime}$}}
\newcommand\arcmin{\hbox{$^\prime$}}

\newcommand\farcs{\hbox{$.\!\!^{\prime\prime}$}}

\newcommand\degr{\hbox{$^\circ$}}
\newcommand*\fs{\ensuremath{\overset{\text{s}}{.}}}
\author{Eleonora Bianchi}
\affiliation{Univ. Grenoble Alpes, CNRS, IPAG, 38000 Grenoble, France}
\email{eleonora.bianchi@univ-grenoble-alpes.fr}
\author{Cecilia Ceccarelli}
\affiliation{Univ. Grenoble Alpes, CNRS, IPAG, 38000 Grenoble, France}
\alsoaffiliation{INAF-Osservatorio Astrofisico di Arcetri, L.go E. Fermi 5, I-50125 Firenze, Italy}
\author{Claudio Codella}
\affiliation{INAF-Osservatorio Astrofisico di Arcetri, L.go E. Fermi 5, I-50125 Firenze, Italy}
\alsoaffiliation{Univ. Grenoble Alpes, CNRS, IPAG, 38000 Grenoble, France}
\author{Juan Enrique-Romero}
\affiliation{Univ. Grenoble Alpes, CNRS, IPAG, 38000 Grenoble, France}
\author{Cecile Favre}
\affiliation{Univ. Grenoble Alpes, CNRS, IPAG, 38000 Grenoble, France}
\author{Bertrand Lefloch}
\affiliation{Univ. Grenoble Alpes, CNRS, IPAG, 38000 Grenoble, France}

\title[ClassI]
{Astrochemistry as a tool to follow the protostellar evolution: the Class I stage}

\setkeys{acs}{keywords = true}
\abbreviations{ISM,\colorbox{yellow}{ASW}}
\keywords{Astrochemistry, low-mass protostars, Class I protostars, protostellar evolution, deuteration, interstellar complex organic molecules}

\mciteErrorOnUnknownfalse
\begin{document}

\begin{abstract}
The latest developments in astrochemistry have shown how some molecular species can be used as a tool to study the early stages of the solar-type star formation process. Among them, the more relevant species are the interstellar complex organic molecules (iCOMs) and the deuterated molecules. Their analysis give us information on the present and past history of protostellar objects. 
Among the protostellar evolutionary stages, Class I protostars represent a perfect laboratory in which to study the initial conditions for the planet formation process. Indeed, from a physical point of view, the Class I stage is the bridge between the Class 0 phase, dominated by the accretion process, and the protoplanetary disk phase, when planets form.
Despite their importance, few observations of Class I protostars exist and very little is known about their chemical content.
In this paper we review the (few) existing observations of iCOMs and deuterated species in Class I protostars.
In addition, we present new observations of deuterated cyanoacetylene and thioformaldehyde towards the Class I protostar SVS13-A.
These new observations allow us to better understand the physical and chemical structure of SVS13-A and compare the cyanoacetylene and thioformaldehyde deuteration with other sources in different evolutionary phases.
\end{abstract}

\section{1. Introduction}

Since the birth of radio-astronomy, a large number of simple and complex molecules have been detected in the galactic Interstellar Medium (ISM).
To date around 200 molecules have been identified mainly by their rotational transition lines \citep{McGuire2018}. 
Among them, molecular species of particular interest are the so-called interstellar complex organic molecules (hereinafter iCOMs; C-bearing molecules
containing at least six atoms \citep{Herbst2009,Ceccarelli2017}) and deuterated molecules (where one or more hydrogen atoms are substituted by deuterium ones). The former, although quite simple from a "terrestrial" chemical point of view, might represent the small bricks from which prebiotic chemistry could start. The latter have a huge diagnostic power in reconstructing the past history of the planetary system forming disk \citep{Ceccarelli2014}.
In our Galaxy, both iCOMs and deuterated species have been detected around low-mass protostars, which are nascent systems that eventually might become similar to our Solar System.

The interest in studying these two classes of molecules is, thus, twofold.
On the one hand, the study of iCOMs can constrain their formation and destruction pathways in a variety of extreme physical conditions dominated by very low temperatures and densities, hugely different from those in terrestrial laboratories. This, in turn, can help to understand possible formation/destruction pathways of even more complex molecules which are impossible to detect in protostellar sources because they have a very low abundance and/or a large partition function, and which might have a role in the emergence of life.
On the other hand, both deuterated and complex organic molecules are powerful diagnostic tools to study the present and the past physical conditions, namely the history of forming/formed systems\citep{Caselli2012, Ceccarelli2014, Bianchi2017a, Bianchi2019}.
In other words, studying the evolution of iCOMs and deuterated molecules in the different stages of star formation could help us to understand how much reprocessing occurs during the birth of a planetary system like our Solar System and which molecules are inherited from the early stages.

In this context, the Class I protostars represent a crucial phase: they are the bridge between the youngest protostars, dominated by the gravitational collapse, and the more evolved planetary disks, where planets and comets form. In addition, as we will discuss in detail in the next section, mounting evidence shows that planets/comets formation occurs very early, likely already in the Class I phase.

In this article, we will review what we know so far about the chemical content of Class I protostars, with a particular focus on iCOMs and deuterated molecules. We will then present new observations towards the prototype Class I source, SVS13-A, which allow us to better understand its structure and its deuterated molecular content.
The structure of this article is the following: in Section 2, we describe why Class I protostars are crucial in the evolution of solar-type systems; in Section 3, we report an overview of the previous observations of iCOMs and deuterated molecules towards Class I protostars; in Section 4, we present new observations of HC$_3$N, DC$_3$N, H$_2$CS and HDCS towards SVS13-A and we discuss their implications in our understanding of Class I protostars; Section 5 concludes providing the future perspectives in the study of Class I protostars.

\section{2. Class I sources, a crucial intermediate evolutionary phase in the formation of solar-type systems}\label{classI-intermediate}

\subsection{2.1 Physical evolution of a solar-type protostar}
\begin{figure}
\begin{center}
\includegraphics[width=7.5cm]{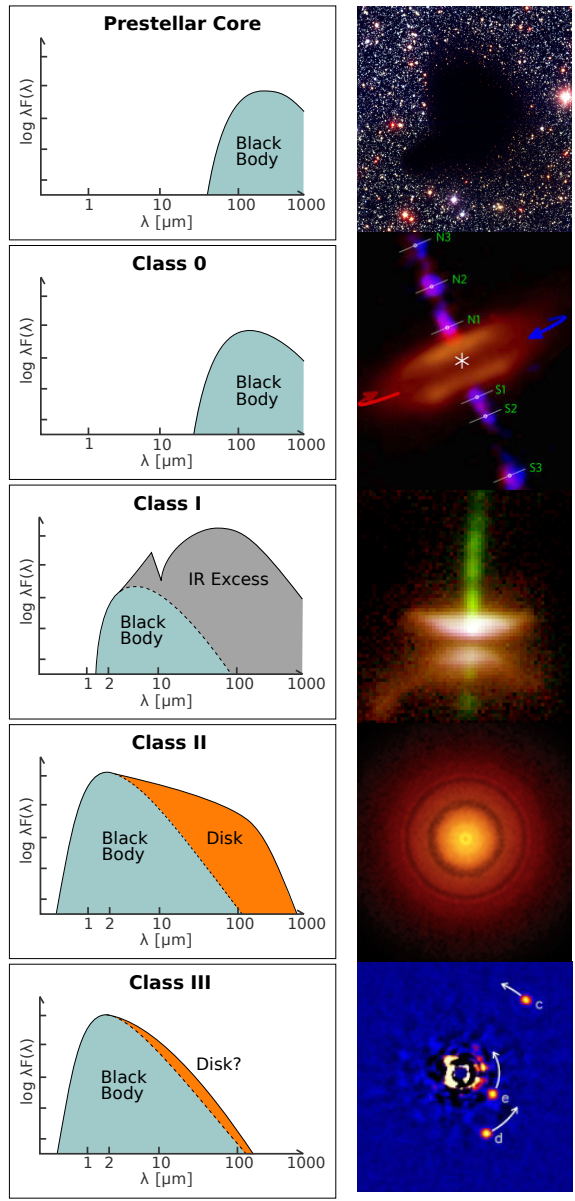}
\caption{{\it Left panels}: Spectral energy distributions (SEDs) of the different formation stages of a solar-type star (adapted from \citep{Persson2014SED}). {\it Right panels}: Real images of objects in the different evolutionary stages taken with different instruments. The prestellar core is the dark cloud B68 as observed by VLT/FORS1 (Credit:ESO). The Class 0 object is the HH212 protostar in Orion as observed by ALMA \citep{Lee2017b}. The Class I object is HH30 as observed by the Hubble Space Telescope (Credit: Chris Burrows (STScI), the WFPC2 Science Team and NASA/ESA). The Class II object is an ALMA view of the protoplanetary disc surrounding the young star TW Hydrae (Credit: S. Andrews (Harvard-Smithsonian CfA); B. Saxton (NRAO/AUI/NSF); ALMA (ESO/NAOJ/NRAO). The Class III object is the image of the system HR 8799 with three orbiting planets. The image has been acquired at the Keck II telescope (Credit: \citep{Marois2010}).} 
\label{Fig:SED}
\end{center}
\end{figure}

The formation of a solar-type star is a complex process which lasts some Myr. It starts in dense filaments of molecular clouds and is governed by the interplay of gravitational collapse, magnetic fields and turbulence \citep{Shu1977,Shu1987,Galli2006,Hennebelle2008,Andre2014,Klessen2016}.
The current paradigm is illustrated in Figure \ref{Fig:SED}, which shows the major phases of the solar-type planetary system evolution, from the prestellar cores to the formation of a planetary system \citep{Dunham2014}. The left panels show the spectral energy distribution (SED) of objects representative of each phase while the right panels show actual images of objects. In the following, we briefly describe each phase.

{\it Prestellar cores:} Inside the filaments of molecular clouds, denser clumps form on scale of $\sim$ 0.1--0.01 pc and some of them, called prestellar cores, slowly accrete matter towards the center under the gravitational force counteracted by the magnetic field. These objects are characterized by high central densities ($n_{H_2}>$ 10$^{4}$ cm$^{-3}$) and low temperatures (around 10 K or less) and are the nursery where one or more protostars are born.

{\it Class 0 protostars:} Once the collapse takes over, the material freely falls towards the center feeding the central object which, together with the infalling envelope, forms the so-called Class 0 protostar. In this stage, the central object is deeply embedded and obscured by the large infalling envelope. 
The Class 0 phase is dominated by the central object accretion through a rotating and accreting disk. The angular momentum is removed thanks to the ejection of gas through high-velocity and highly collimated protostellar jets.

{\it Class I protostars:} After about 10$^{5}$ yr, the protostar reaches the Class I phase in which the large-scale envelope is partially swept up. The central regions of the protostars start to be visible in the protostar SED, which is characterized by an infrared excess due to the central object light absorbed and scattered by the dusty disk and envelope.

{\it Class II and III protostars:} After about 10$^{6}$ yr, the protostar enters the Class II phase, when the envelope is almost entirely dissipated. The accretion disk becomes a protoplanetary disk, where the dust coagulation process becomes more and more efficient. The system ends up with the formation of a debris disk during the Class III phase and a final planetary system. 

Thanks to observations performed at 0.96--3.8 $\mu$m with the SPHERE instrument, a young planet has been recently directly imaged towards the PDS 70 transition disk\citep{Muller2018}. However, at present it remains difficult to directly detect planet(s) in formation (likely because of instrumental limits).
While ongoing planet formation is a plausible origin of sub-structure in protoplanetary disks, it is not the only one. Many other physical processes such as large scale vortices, magnetic reconnection, and condensation fronts, have been invoked (e.g.\citep{Suriano2018,Barge2017,Ruge2016, Zhang2015, Gonzalez2015, Flock2015}).
Nonetheless, hydrodynamical simulations predict that a planet in formation will open gaps in both dust and gas (e.g.\citep{Oberg2015b, Baruteau2016, Facchini2018}). Therefore, it remains possible to indirectly observe a forming planet through observations of the induced planet-(gas and dusty) disk interaction. In that light, ALMA observations showed that rings and gaps structure, in dust at least, seems to be a common feature in Class II protoplanetary disks of different ages (e.g. \citep{Andrews2016,Isella2018,Fedele2017,Loomis2017,Andrews2018, Dipierro2018,vanderMarel2019}). In addition, investigations of gas perturbations and/or gas kinematics motions towards the protoplanetary disks surrounding the T-Tauri star AS 209 and the Herbig object HD 163296 strongly suggest the presence of one or more embedded forming planets are present in these objects\citep{Favre2019, Teague2018, Pinte2018}.

The spectacular images provided by ALMA on protoplanetary disks suggest that planet formation could begin already at the Class I stage, earlier than previously thought.
Indeed, the recent observations of the HL Tau system \citep{ALMA2015}
showed several sharp rings and gaps in the disk which could indicate 
the presence of nascent planets \citep{Dong2015}.
The youth of this system, thought to be $\le$ 1 Myr old, in a transition stage 
between Class I and II, indicates that planet formation could start during the early Class I stage.
Several studies have been dedicated to the measurement of the disk masses in Class I
sources \citep{Evans2009, Sheehan2017} with the intent to determine the
initial mass budget for forming planets.
Recent studies in fact indicate that the protostellar disk mass is already set near 
the onset of the Class 0 protostellar stage and remains roughly constant during the Class I
protostellar stage \citep{Andersen2019}.

 In summary, the Class I protostars promise to be the best targets to study the initial conditions of planet formation, especially from the chemical point of view, since their envelope+disk system is still massive enough to be detected and, hence, studied with millimeter telescopes and interferometers.

\subsection{2.2 Evolution of the iCOMs content}

The different evolutionary phases described above correspond to different degrees of complexity of matter chemical composition (see e.g. \citep{Caselli2012, Ceccarelli2014, Yamamoto2017}). In the following, we briefly describe the evolution of iCOMs in the different phases.

{\it Prestellar cores:} At the low temperatures found in molecular clouds and prestellar cores, atoms and molecules in the gas phase freeze-out onto the cold surfaces of the dust grains. Hydrogenation of these frozen atoms and molecules takes place on the grain surfaces, forming icy mantles, mostly composed of water (H$_{2}$O). Relevant to this article, the frozen CO is hydrogenated and forms formaldehyde (H$_{2}$CO) and methanol (CH$_3$OH) \citep{Watanabe2002, Rimola2014}. These simple hydrogenated species are detected in the solid-phase via infrared observations \citep{Boogert2015}.
More complex species, notably some iCOMs such as methyl formate (HCOOCH$_{3}$) and dimethyl formate (CH$_3$OCH$_3$), are detected in the gas phase towards prestellar cores \citep{Oberg2010,Bacmann2012,Cernicharo2012,Vastel2014,Jimenez2016,Punanova2018}. It is not clear yet how these iCOMs are synthesised. In principle, at these low temperatures, no grain surface reactions other than hydrogenation can occur nor can these large molecules be released from the grain surfaces into the gas-phase. Thus, it is proposed that they form via gas-phase reactions from simple species like methanol, injected from the grain mantles by non-thermal desorption processes \citep{Vasyunin2013,Balucani2015}.

{\it Class 0 protostars:} During the Class 0 phase, the forming central object increasing luminosity warms up the surrounding envelope. Where the temperature reaches about 100 K, the grain mantles sublimate, giving origin to the so-called hot corinos \citep{Ceccarelli2017,Caselli2012}. In these objects, a plethora of iCOMs (more than a dozen \citep{Ceccarelli2017}) are detected in relatively large abundances, similar to the methanol one \citep{Cazaux2003,Taquet2015,Jorgensen2016,Calcutt2018}.
The synthesis of iCOMs in hot corinos is debated. One possibility is that, during the warming up of the envelope, atoms and radicals, previously formed and frozen in the mantles, acquire mobility and react on the grain surfaces forming iCOMs \citep{Garrod2006, Garrod2008}. Then, at about 100 K the entire mantle sublimates injecting into the gas-phase the iCOMs formed in this way. Theoretical chemistry studies, however, show that this combination of radicals does not necessarily end up in iCOMs \citep{Rimola2018,Enrique-romero2019}.
Alternatively, the mantle simple molecules (such as methanol or formaldehyde) formed during the prestellar phase are injected into the gas at 100 K and undergo reactions that form iCOMs \citep{Skouteris2017,Skouteris2018,Skouteris2019}.

{\it Class I protostars:} Section 3.1 will review the status of the art, but we anticipate that, unfortunately, very little is known about the iCOMs census and abundance in Class I protostars. 

{\it Protoplanetary disks:} In protoplanetary disks, very few iCOMs have been detected so far: methanol (CH$_3$OH), methyl cyanide (CH$_3$CN) and formic acid (HCOOH) \citep{Oberg2015a,Bergner2018,Favre2018}. The obvious question is: are more complex molecules not detected in protoplanetary disks only because of the current instrument detection limits or chemistry changes from the Class 0 to the protoplanetary disk phase? 
The fact that the existing detections are all at the 5$\sigma$ letter definitely points to instrumental limits, and further study of iCOMs content in Class I protostars might answer this question.

In summary, Class I sources are, from an evolutionary point of view, the connection between the
Class 0 protostars and the protoplanetary disks (see previous Section). However, it remains unclear whether this is also true from a chemical point of view.
Analogously, the degree of reprocessing of the chemical content from the earlier to the late evolutionary stages is still poorly constrained. Some similarities observed in the molecular abundances of Solar System comets (e.g. 67P/Churyumov-Gerasimenko) and young protostars \citep{LeRoy2015,Droz2018,Bianchi2019}
seem to suggest that something could survive the journey.
The interest for Class I protostars is even higher when we think that the nascent planetary atmospheres could be linked to the early disk chemical composition.

\subsection{2.3 Molecular deuteration as an evolutionary indicator}
The process of deuteration consists of an enrichment of the amount of deuterium with respect to hydrogen in molecular species.
Despite the very low level of the D/H elementary abundance ($\sim$ 1.6 10$^{-5}$)\citep{Linsky2006}, molecules in cold ISM show a high ratio, up to 13 orders of magnitude larger than the elemental one\citep{Parise2004}. 

\begin{figure}
\begin{center}
\includegraphics[angle=0,width=18cm]{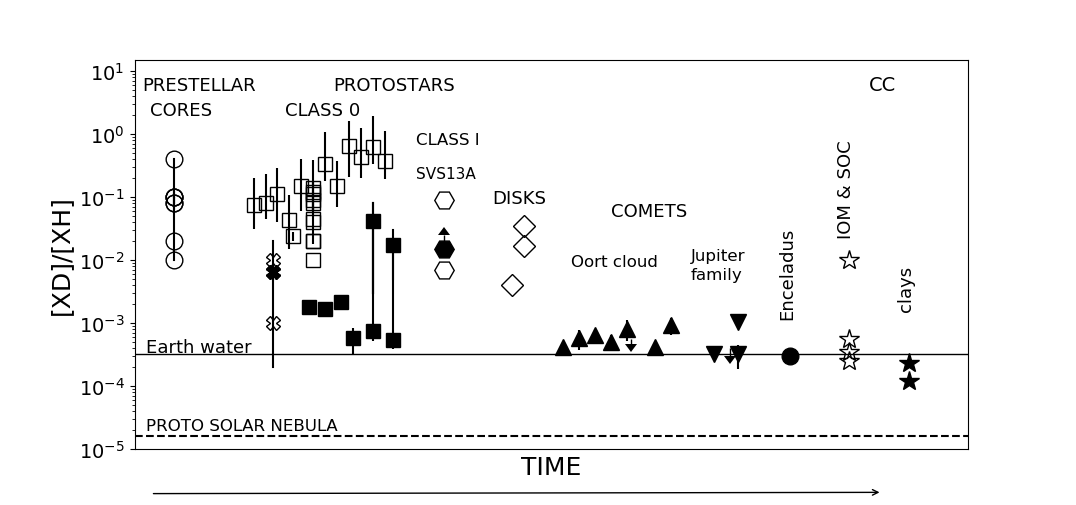} 
 \caption{Molecular deuteration ratio of ISM and Solar System objects (adapted from \citep{Ceccarelli2014}). The filled symbols refer to water whereas open symbols refer to organic matter. Note that for organic matter, different molecular tracers are used.
 The measurements of prestellar cores deuteration, indicate with circles are from \citep{Spezzano2013, Bizzocchi2014, Chacon2019}. Class 0 protostars are indicated with squares and are from \citep{Parise2006, Taquet2013, Coutens2012,Persson2014, Jensen2019}.
 The previous data have been updated with the last measurements at a Solar System scale provided by the ALMA PILS survey \citep{Coutens2016, Persson2018, Jorgensen2018} and HH212 \citep{Bianchi2017b}. 
The crosses indicate deuteration measurements in the protostellar shock region L1157-B1\citep{Codella2012, Fontani2014}.
 The Class I measurements (hexagons) refer to SVS13-A, specifically to the following molecular species: formaldehyde\citep{Bianchi2017a}, methanol\citep{Bianchi2017a} and water \citep{Codella2016b}. The diamond symbols refer to deuteration measurements in protoplanetary disks\citep{vanDish2003, Guilloteau2006, Oberg2012}.}
  \label{Fig:Deut}
  \end{center}
\end{figure}

Deuterated molecules have been observed in all the stages of the solar-type star formation process, from prestellar cores \citep{Bacmann2003,Vastel2003,Vastel2004,Vastel2006,Bizzocchi2014,Chacon2019}, to Class 0 \citep{Ceccarelli2001,Parise2002,Parise2004,Parise2006,Coutens2016,Manigand2019} and protoplanetary disks \citep{vanDish2003,Guilloteau2006,Oberg2012,Huang2017,Teague2015,Salinas2017}. These observations are a powerful tracer of the evolution during the formation of a solar-type planetary system \citep{Ceccarelli2014}. 
Fig. \ref{Fig:Deut} shows the deuteration measured for water and organic molecules at different stages of the solar-type star formation process, from prestellar cores to Solar System objects. 
The figure suggests a general decreasing trend which can be interpreted as a modification of the gas chemical content.
In particular, the deuteration of organics, considering different molecular tracers, decreases up to two orders of magnitude going from Class 0 protostars to the Solar System objects such as comets and carbonaceous chondrites \citep{Ceccarelli2014}.
However, measurements of deuteration in protoplanetary disks are based on the detection of DCN and DCO$^+$ in few objects \citep{Guilloteau2006, Huang2017, Mathews2013, Oberg2010, Oberg2011disks, Oberg2015a, Qi2008, Qi2015, Salinas2017, Teague2015, vanDish2003} . 
Moreover, in these objects the emission is probably tracing the disk surface, due to substantial disk midplane freeze-out. Given the difficulty so far to observe other large deuterated molecules in disks that trace more closely to the midplane, it would be very useful to understand if changes in deuteration occur already during the protostellar stages.

Briefly, the molecular deuteration is enhanced in cold environments and it is mostly, but not exclusively, caused by the enhancement of the H$_2$D$^+$ abundance with respect to the H$_3^+$ one. This is the molecule that, in molecular gas, transfers the deuterium from HD to the other molecules (see \citep{Ceccarelli2014} for more details). The major parameters that influence the molecular deuteration are gas temperature \citep{Watson1974}, CO abundance in the gas phase \citep{Roberts2000,Bacmann2003} and H$_2$ ortho-to-para ratio \citep{Flower2006,Bovino2017}.

As mentioned above, in protostellar objects the measured degree of deuteration can be up to 13 orders of magnitude larger than the elementary value and it is different for each molecular species \citep{Caselli2012}. 
In general, the external region of the protostellar envelopes (at larger radii than 100 au) shows gas conditions very similar to that observed in prestellar cores. The measured molecular deuteration in these regions is therefore very likely representative of the current gas conditions.
On the contrary, the molecular content in hot corinos is dominated by the ice mantle sublimation (see above). Therefore, molecular deuteration measured in these regions is dominated by deuteration at the time of mantle formation and, consequently, can give us information on past chemical gas conditions and the history of the mantle formation \citep{Ceccarelli2001,Taquet2012b,Codella2012,Caselli2012,Ceccarelli2014}. 
Water, for example, that is mostly formed in molecular clouds, presents a lower degree of deuteration with respect to formaldehyde and methanol, which are very likely formed during the prestellar phase.
Interestingly, based on the deuteration measurement, recent modeling studies suggest that interstellar water ice is largely inherited by forming disks without significant alteration \citep{Cleeves2014,Furuya2016, Furuya2017}.

\section{3. iCOMs and molecular deuteration in Class I protostars}
In this Section, we review what is known, namely observed, about iCOMs and molecular deuteration in Class I protostars.

\subsection{3.1 iCOMs}

Several observations have been carried out so far with single-dish telescopes in the mm-spectral range to detect iCOMs in samples of Class 0 and Class I protostars (see e.g. \citep{Caux2011, Bottinelli2004, Maret2005, Bottinelli2007, Oberg2011, Oberg2014, Graninger2016, Lefloch2018}).
However, no clear results have been obtained
so far regarding the chemical content modification during the evolutionary transition from Class 0 to Class I, because of the lack of surveys dedicated to Class I protostars.
This is also related to the instrumental limitation of single-dish telescopes, which provide a relatively small angular resolution and large beam dilution, severely affecting the observations of compact and weak sources such as Class I protostars.
For example, some Class I protostars have been included in a survey performed using the IRAM-30m with angular resolution larger than 20$\arcsec$ \citep{Oberg2011, Oberg2014, Bergner2017, Law2018, Lefloch2018}. As a result, iCOMs are almost non-detected or detected via very low
excitation lines. Moreover, these surveys often derive low rotational temperatures, which suggests that the detected iCOMs lines might not originate in the hot corinos.

SVS13-A has been so far the only Class I source for which a complete census of the iCOMs (with abundances larger than about $10^{-9}$) has been possible \citep{Bianchi2019}. The source will be described in detail in the next Section; here we anticipate that it has a hot corino surrounded by a cold envelope, namely a structure similar to a Class 0 protostars, but with a less massive envelope.

\begin{figure*}
\begin{center}
\includegraphics[width=15.5cm]{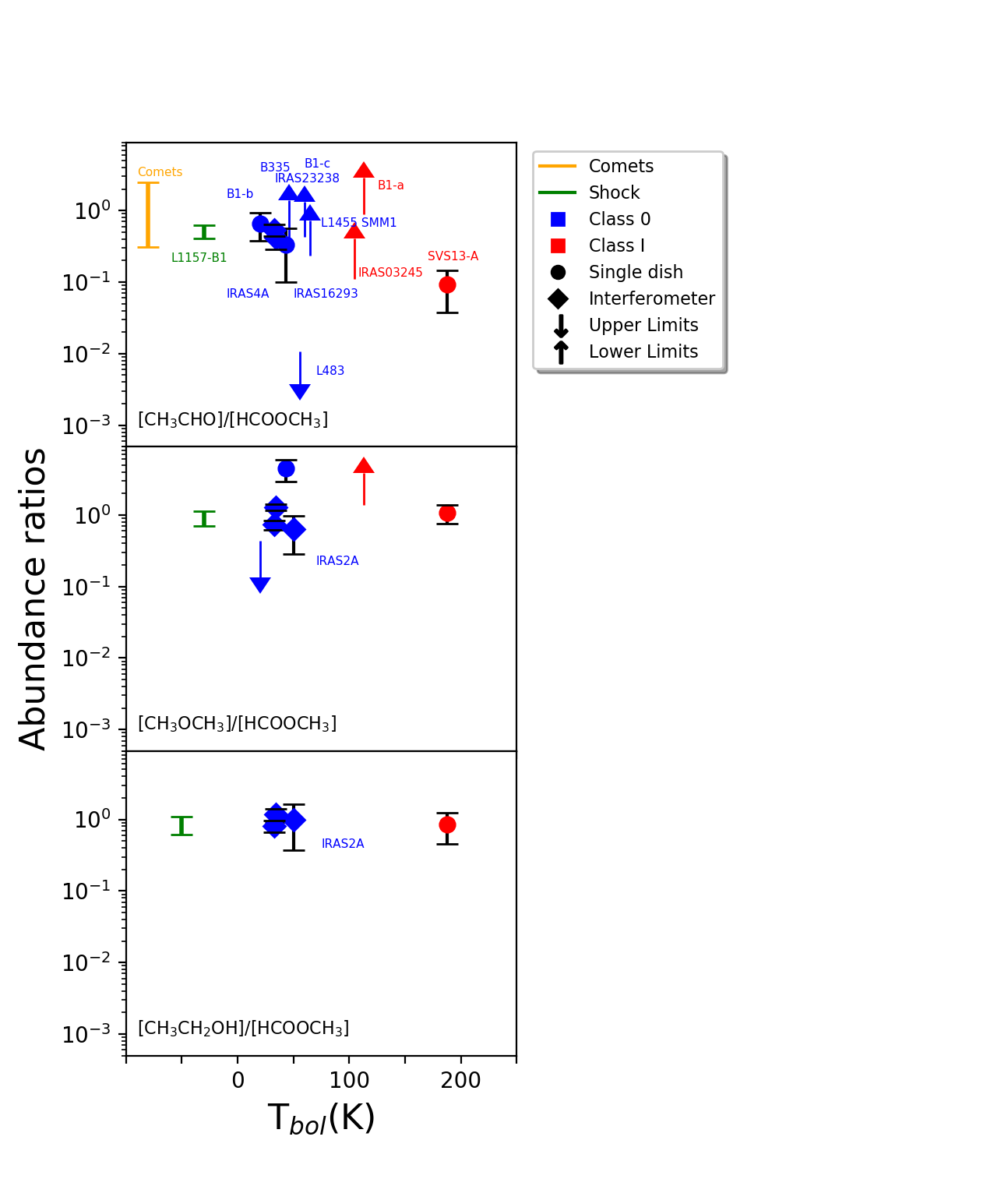}
\caption{Figure adapted from \citep{Bianchi2019}. Abundance ratios of the iCOMs detected in SVS13-A compared to different sources, as indicated in the upper panel. The sources are: the pre-protostar hydrostatic core B1-b\citep{Pezzuto2012, Gerin2015}; the Class 0 sources IRAS4A, IRAS2A\citep{Taquet2015, Lopez2017}, IRAS16293--2422 (e.g. \citep{Jaber2014, Jorgensen2016, Jorgensen2018}), B1-c\citep{Oberg2014, Bergner2017}, IRAS23238+7401 \citep{Graninger2016, Bergner2017}, L1455 SMM1 \citep{Graninger2016, Bergner2017}, IRAS19347+0727 in B335 \citep{Imai2016} and L483 \citep{Oya2017}; the protostellar shock L1157-B1 \citep{Codella2010, Lefloch2017}, the Class I sources B1-a \citep{Oberg2014, Bergner2017}, IRAS03245+3002 \citep{Graninger2016, Bergner2017} and SVS13-A \citep{Bianchi2019}; comets\citep{LeRoy2015}. Blue symbols indicate Class 0 protostars while red symbols are for Class I protostars. Circles indicate single-dish measurements while diamonds are for interferometric measurements. Arrows indicate upper limit measurements. The abundance ratios of comets and the protostellar shock L1157-B1 are reported for comparison using an orange line and a green line, respectively. Note that in these cases the x-value has no meaning.}
\label{Fig:Total}
\end{center}
\end{figure*}

The iCOMs census towards SVS13-A was obtained in the framework of the IRAM-30m Large Program ASAI (Astrochemical Surveys At IRAM: Lefloch et al.\citep{Lefloch2018}), which obtained unbiased spectral surveys of several solar-type protostars, entirely covering the 1, 2 and 3 mm wavelength bands observable from ground. Given the small size of the SVS13-A hot corino ($\lesssim 1''$), iCOMs were mostly detected in the 1mm band, less affected by the beam dilution. The large spectral range covered by the survey allowed us to detect more than 100 lines from different iCOMs and, consequently, to accurately determine the lines excitation conditions and the iCOMs column densities. In particular, we derived rotational temperatures between 35 and 110 K, and column densities between 3 $\times$ 10$^{15}$ and 1 $\times$ 10$^{17}$ cm$^{−2}$ on the 0$\farcs$3 size previously determined by interferometric observations of glycolaldehyde\citep{Desimone2017}.
Figure \ref{Fig:Total} shows some of the measured abundance ratios towards SVS13-A. The abundance ratio of some key iCOMs such as acetaldehyde, dimethyl ether and ethanol with respect to methyl formate are compared for Class 0 sources and SVS13-A and two other Class I sources, whose values are from the literature \citep{Oberg2014, Graninger2016, Bergner2017}. These species are of particular interest because they are abundant and they can give us information on different formation mechanisms. For comparison, the figure also reports the abundance ratios measured in comets \citep{LeRoy2015} and in the molecular-rich protostellar shock L1157-B1 \citep{Codella2010,Lefloch2017}.
The iCOMs abundances are normalized to methyl formate to avoid opacity problems which could affect other more abundant species such as methanol.
Strictly speaking, this figure suggests that when iCOMs are detected in Class I sources, the iCOM composition appears similar to what is seen at the Class 0 stage. More loosely, the chemical richness of Class I protostars is comparable to that of Class 0 sources.
Therefore, the chemical complexity could be inherited from the previous stages, although the comparison is practically based on only one Class I source, SVS13-A, which may not be representative of the whole Class I. For the other two Class I sources, the measurement are mostly only upper limits. A larger sample of source is needed to confirm this picture.

Recently, two Class I sources from the Serpens cluster were observed using ALMA interferometer \citep{Bergner2019}. Only one on them, the source Ser-emb 17, shows iCOMs emission from CH$_{\rm 3}$OH, CH$_{\rm 3}$OCH$_{\rm 3}$, HCOOCH$_{\rm 3}$, NH$_{\rm 2}$CHO and CH$_{\rm 2}$CO.
The abundance ratios of iCOMs with respect to methanol in this source are comparable to that measured in the other two Class 0 Serpens sources contained in the sample, Ser-emb 1 and Ser-emb 2, confirming that suggested by the analysis of SVS13-A.

\subsection{3.2 Molecular deuteration}

As mentioned in Section 2.3, deuterated molecules have been detected in all the stages of Sun-like star formation,
from the prestellar core stage to the Solar System objects \citep{Ceccarelli2014} but no clear results have been obtained so far for the intermediate Class I evolutionary phase.
The first indications a possible decrease of the deuteration with the evolutionary stage came from measurements of the double deuterated formaldehyde in star-forming regions \citep{Loinard2002}.
Other measurements of deuteration in Class I sources \citep{Roberts2007} included only few transitions and sampled large regions (up to 58$\arcsec$), well beyond the protostellar system.
In addition, a low deuteration of formaldehyde was reported towards R CrA IRS7B, a low-mass protostar in the Class 0/I transitional stage \citep{Watanabe2012}. However, in this case the envelope chemical composition is probably altered by the external UV irradiation from the nearby Herbig Ae star R CrA so that the low deuteration ratio cannot be interpreted as an evolutionary trend.
\begin{figure}
\begin{center}
\includegraphics[angle=0,width=14cm]{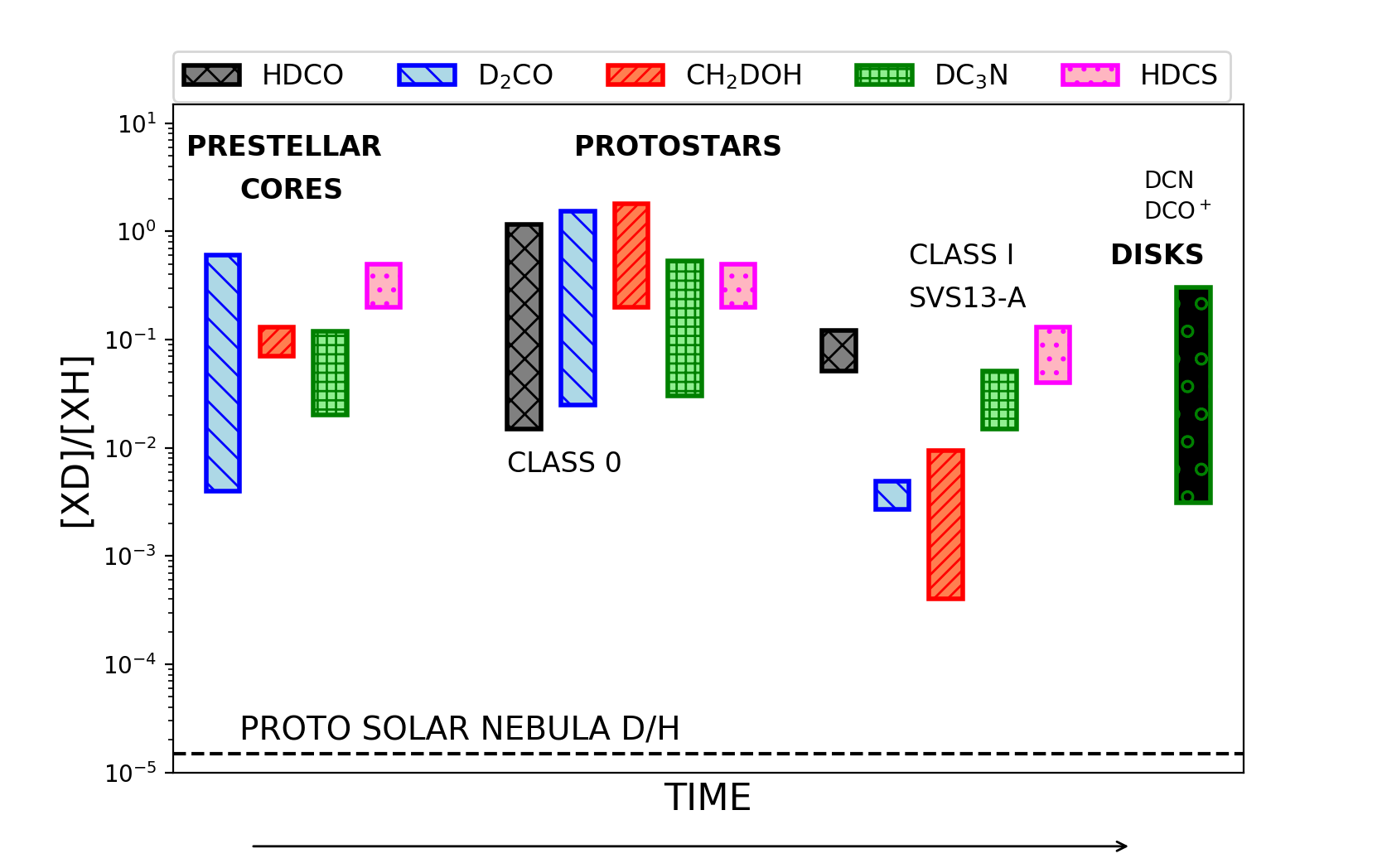} 
  \caption{[XD]/[XH] ratio measured in organic matter in different astronomical sources: prestellar cores (D$_{\rm 2}$CO\citep{Bacmann2003}, CH$_{\rm 2}$DOH\citep{Bizzocchi2014}, DC$_{\rm 3}$N\citep{Langer1980, Howe1994, Cordiner2012}, and HDCS\citep{Marcelino2005,Vastel2018}), Class 0 protostars (HDCO \citep{Parise2006}, D$_{\rm 2}$CO\citep{Parise2006}, CH$_{\rm 2}$DOH \citep{Parise2006}, DC$_{\rm 3}$N\citep{Sakai2009b}, and HDCS\citep{Droz2018}) .
SVS13-A data refer to the deuteration of HDCO (8.6 $\times$ 10$^{-2}$), D$_{\rm 2}$CO (3.8 $\times$ 10$^{-3}$)\citep{Bianchi2017a}, CH$_{\rm 2}$DOH (7.1 $\times$ 10$^{-3}$ for the hot corino and 1.5 $\times$ 10$^{-3}$ for a larger region, i.e. a radius $\leq$ 350 au)\citep{Bianchi2017a}, DC$_{\rm 3}$N (1.5 -- 3.6 $\times$ 10$^{-2}$, this work), and HDCS (4 -- 9 $\times$ 10$^{-2}$, this work). 
Protoplanetary disks data refer to measurements of DCN/HCN and DCO$^{+}$/HCO$^{+}$ \citep{Guilloteau2006, Huang2017, Teague2015, vanDish2003, Qi2008, Mathews2013, Oberg2010, Salinas2017}.
Figure adapted from \citep{Bianchi2017a}.
We have to note that opacity can strongly affect the main isotopologues abundances and then the [XD]/[XH] derivation (see text).}
  \label{Fig:Deut_ele}
  \end{center}
\end{figure}

In SVS13-A, the singly deuterated formaldehyde (HDCO) is $\sim$0.09 with respect to formaldehyde\citep{Bianchi2017a}, a value similar to the average one measured in Class 0 protostars ($\sim$0.12\citep{Parise2006}), as shown in Fig. \ref{Fig:Deut_ele}. Please note that this comparison only refers to Class 0 protostars observed with the same single-dish telescope (IRAM-30m): measurements obtained with higher resolution (e.g. with ALMA) show lower levels of deuteration, which may be due to the different probed spatial scales, namely the fact that multiple physical components are present in the line of sight. Therefore, we emphasize that comparisons based on single-dish data have to be taken with caution.

With this caution in mind, the deuteration measured using doubly deuterated formaldehyde (D$_2$CO) and singly deuterated methanol (CH$_2$DOH) is one and two orders of magnitude lower, respectively, than that measured towards Class 0 protostars \citep{Bianchi2017a,Parise2006} (see Figure \ref{Fig:Deut_ele}).
We have to caution, though, that methanol deuteration in Class 0 sources suffers from uncertainty in the opacity of the lines used to derive CH$_3$OH column density \citep{Parise2006}. For example, the new analysis of the Class 0 source IRAS4A, obtained with ASAI and that includes $^{13}$CH$_3$OH lines, shows that the CH$_3$OH column density is underestimated by about a factor of ten if not corrected for the opacity (Lefloch, private communication). When considering all these uncertainties, it is not totally clear whether the molecular deuteration is preserved from Class 0 to Class I. 
In summary, accurate opacity estimations together with more observations of both Class 0 and Class I sources are needed to draw firm conclusions.

\section{4. New tracers of molecular deuteration towards the Class I protostar SVS13-A}

In this section, we present new observations of HC$_3$N and H$_2$CS and their deuterated counterparts towards the Class I protostar SVS13-A. Before reporting the new observations and their analysis, we briefly describe SVS13-A, which is the prototype Class I protostar where most of the chemical studies have been carried out so far.

\subsection{4.1 Source background}

SVS13-A is a Class I protostar belonging to the SVS13 cluster, which is composed of at least five objects. It is part of the NGC 1333 molecular cloud in Perseus, and its distance has been accurately measured by GAIA to be 299 $\pm$ 14 pc \citep{Zucker2018}.

Since the discovery of the first dark clouds by Barnard in 1913, the Perseus region has received increasing attention and has been extensively investigated at many wavelengths and with a variety of spatial resolutions.
In particular, the young cluster NGC 1333 is one of the best studied and the most active region of star formation in the Perseus cloud complex. The region is rich in sub-mm cores, embedded Young Stellar Objects (YSOs), radio continuum sources, masers, IRAS sources, SiO molecular jets, H$_2$ and Herbig-Haro shocks, molecular outflows, and the lobes of extinct outflows\citep{Reipurth1993,Lefloch1998a,Lefloch1998b,Codella1999,Tobin2016}. 
Far Infrared and sub-millimeter observations of the southern region revealed many bright Class 0/I protostars, including the famous ones, SVS13, IRAS2 and IRAS4. Successive high spatial resolution observations showed that many of them are actually binary and even multiple systems. 
Further molecular line observations of the region highlighted the importance of protostellar outflows in the matter distribution of NGC 1333.
As an example, Lefloch et al. \citep{Lefloch1998a} showed the presence of two large cavities excavated by the outflows ejected from newly formed stars.

\begin{figure}
\begin{center}
\includegraphics[angle=0,width=12cm]{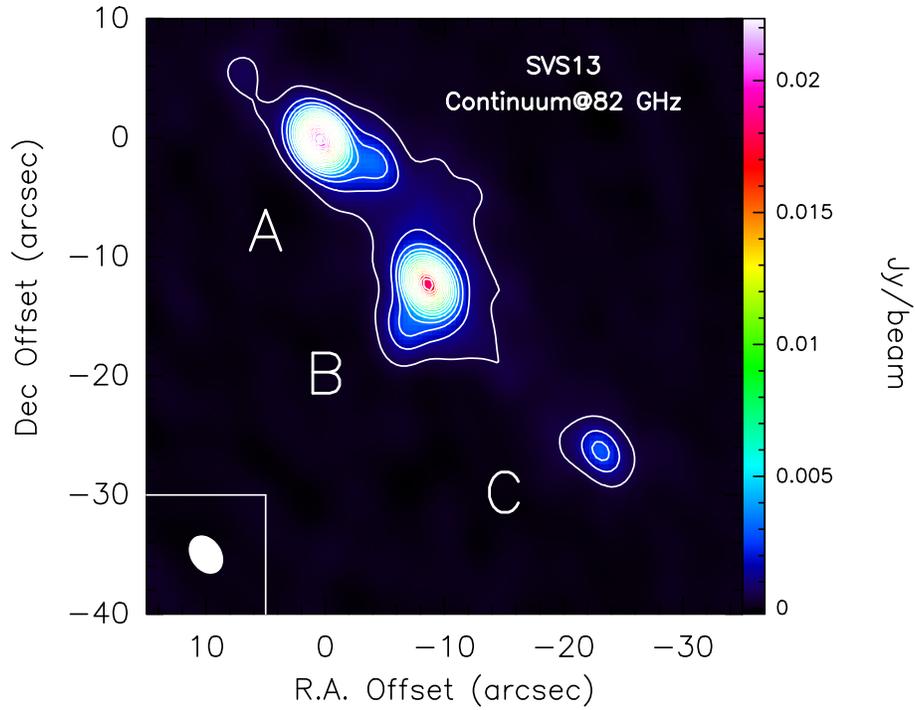} 
  \caption{Continuum observations at 82 GHz of the SVS13 system performed using the IRAM-NOEMA interferometer. The observations are performed in the framework of the Large Program SOLIS (Seeds Of Life In Space \citep{Ceccarelli2017}). The beam size is 3.5$\arcsec$ $\times$ 2.6$\arcsec$ while the position angle (PA) is 31$^\circ$. The SVS13 system is composed by 3 main sources. SVS13-A, in the upper left, is the brighter source in the millimeter and it is classified as a Class I protostar (see text). It has been recently resolved into a close binary system, separated by 0$\farcs$3. SVS13-B is a younger Class 0 source, located $\sim$ 15$\arcsec$ south-east with respect to SVS13-A. Finally, in the lower right of the map there is a third source, SVS13-C which is located at a distance of 20$\arcsec$ from A.}
  \label{Fig:solis}
  \end{center}
\end{figure}

SVS13-A itself is associated with an extended outflow ($>$ 0.07 pc \citep{Lefloch1998a, Codella1999})
 as well as with the famous chain of Herbig-Haro (HH) objects 7--11 \citep{Reipurth1993}.
IRAM-30m observations revealed that the SVS13 star forming region is associated with a YSOs cluster dominated in the millimeter by two objects, called A and B, which are separated by $\sim$ 15$\arcsec$ \citep{Grossman1987,Chini1997}. A third source, SVS13-C, is located farther away at a distance of 20$\arcsec$ from A \citep{Chini1997}.
At smaller scales ($\leq$100 au), the region is even more complex than previously thought, as SVS13-A is itself a binary system, separated by 0$\farcs$3 (30 au) \citep{Rodriguez1999, Anglada2000} (see Figure \ref{Fig:solis}; Bianchi et al. in preparation).

Interestingly, SVS13-A and SVS13-B seem to be in two different evolutionary phases. SVS13-B is a bona fide Class 0 protostar of $L_{\rm bol}$ $\simeq$ 1.0 $L_{\rm sun}$ \citep{Chen2009, Tobin2016} which drives a well collimated SiO jet \citep{Bachiller1998, Lefloch1998b}.
SVS13-A definitely has a larger luminosity, ($L_{\rm bol}$ $\simeq$ 32.5 $L_{\rm sun}$ \citep{Tobin2016}), a low $L_{\rm submm}$/$L_{\rm bol}$ ratio ($\sim$ 0.8\%) and a high bolometric temperature ($T_{\rm bol}$ $\sim$ 188 K, \citep{Tobin2016}). Thus, although still deeply embedded in a large scale envelope
\citep{Lefloch1998a}, SVS13-A is considered a Class I protostar. 

A systematic study of the molecular lines towards SVS13-A has revealed the presence of a hot corino region via the detection of warm ($\geq$150 K) and very compact ($\sim25$ au in diameter) HDO emission \citep{Codella2016b}. 
Later, De Simone et al. \citep{Desimone2017} imaged the line emission from a iCOM, glycolaldehyde, around SVS13-A confirming the existence of a hot corino with size of 0$\farcs$3, corresponding to $\sim$ 90 au in diameter. 
More studies then aimed to better characterize the properties of the SVS13-A hot corino. Bianchi et al. \citep{Bianchi2017a} showed that methanol is very abundant in the hot corino and measured its gas temperature to be around 100 K (as expected).
The SVS13-A hot corino is rich in iCOMs, showing emission from the classical ones, namely acetaldehyde, methyl formate, dimethyl ether, ethanol, and formamide \citep{Bianchi2019}.
Finally, a formaldehyde multiple-line analysis showed the presence of a cold ($\sim$20 K) and relatively extended ($\sim$1500 au in diameter) envelope surrounding the SVS13-A hot corino \citep{Bianchi2017a}.

When it comes to the measurement of the molecular deuteration in SVS13-A, only that of methanol and formaldehyde has been measured before the present work.
The CH$_2$DOH/CH$_3$OH abundance ratio in the hot corino is 0.007$\pm$0.002 whereas in the extended envelope the HDCO/H$_2$CO and D$_2$CO/H$_2$CO abundance ratios are 0.09$\pm$0.04 and 0.004$\pm$0.001, respectively \citep{Bianchi2017a}. 

\subsection{4.2 Observations and results}

We report new observations obtained within the IRAM-30m Large Project ASAI \citep{Lefloch2018}.
The data were acquired during several runs between 2012 and 2014 using the broad band EMIR receiver, connected to FTS200 backends, with a spectral resolution of 200 kHz.
The rms noise (in T$_{\rm MB}$ scale) is about 2 mK, 7 mK, 9 mK in a channel of 0.6 km s$^{-1}$, 0.4 km s$^{-1}$, and 0.2 km s$^{-1}$ for the 3, 2, and 1 mm spectral windows, respectively. The observations pointed towards SVS13-A, namely at $\alpha_{\rm J2000}$ = 03$^{\rm h}$ 29$^{\rm m}$
03$\fs$76, $\delta_{\rm J2000}$ = +31$\degr$ 16$\arcmin$ 03$\farcs$0, and they were acquired in wobbler switching mode (with a 180$\arcsec$ throw). 
The pointing was found to be accurate to within 3$\arcsec$. The telescope HPBWs lies from $\simeq$ 9$\arcsec$ at 276 GHz to $\simeq$
30$\arcsec$ at 80 GHz. 

Data reduction and lines Gaussian fit have been performed using the GILDAS-CLASS\footnote{www.iram.fr/IRAMFR/GILDAS/} package. Besides the errors on the Gaussian fit, the calibration uncertainties are stimated to be $\simeq$ 20$\%$. For a more
detailed description of the ASAI observations we refer the reader to \citep{Lefloch2018}.

\subsubsection{4.2.1 Cyanoacetylene}
We detected 13 and 7 lines from cyanoacetylene and its deuterated form DC$_{\rm 3}$N, respectively.
The excitation energies,$E_{\rm up}$, cover the range 20--165 K and 22--69 for HC$_{\rm 3}$N and DC$_{\rm 3}$N respectively.

\begin{figure}
\begin{center}
\includegraphics[angle=0,width=10cm]{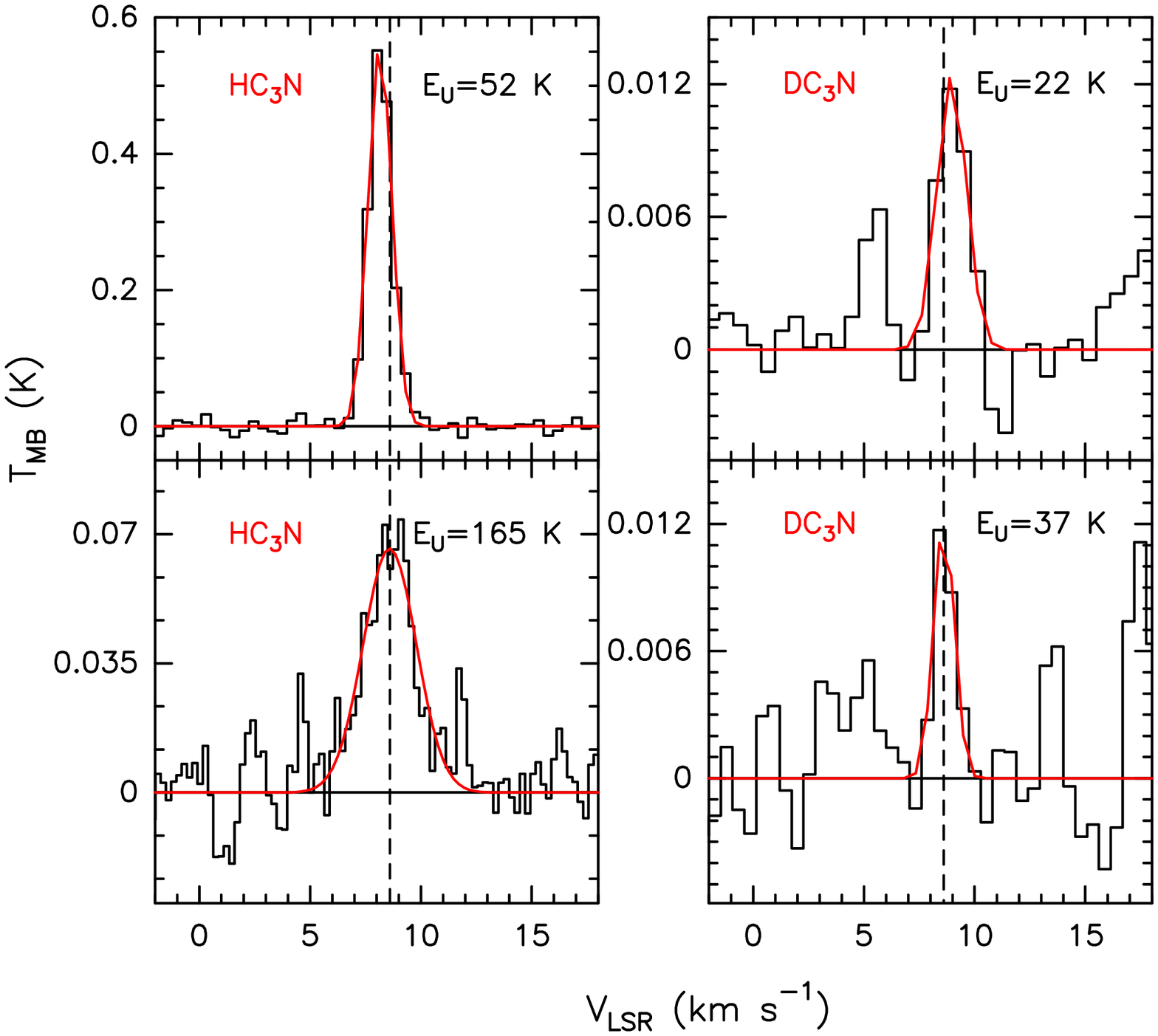} 
  \caption{Examples of the observed spectra of HC$_{\rm 3}$N and DC$_{\rm 3}$N, in T$_{\rm MB}$ scale (i.e. not corrected for the beam dilution): species and excitation energies are reported in the upper left and right corner of each panel, respectively. The vertical dashed line stands for the ambient LSR velocity (+ 8.6 km s$^{-1}$)\citep{Chen2009}.}
  \label{Fig:spectra-HC3N}
  \end{center}
\end{figure}

Examples of the detected line spectra are shown in Figure \ref{Fig:spectra-HC3N}. Table 1 reports the detected transitions and the observational parameters. 
The peak velocities of the detected HC$_{\rm 3}$N and DC$_{\rm 3}$N lines are between +7.7 and +9.0 km s$^{-1}$, being consistent, once the fit and the calibration uncertainties are considered, with the systemic source velocity.
The line profiles are close to a Gaussian shape.
The line FWHMs are between 1.1 and 2.8 km s$^{-1}$ for the main isotopologue and between 0.4 and 1.9 km s$^{-1}$ for the deuterated isotopologue.

\begin{table*}
\resizebox{\textwidth}{!}{
\caption{List of transitions and line properties (in $T_{\rm MB}$ scale) of the DC$_{\rm 3}$N and HC$_{\rm 3}$N emission detected towards SVS13-A.}
\begin{tabular}{lccccccccc}
 \hline
\multicolumn{1}{c}{Transition} &
\multicolumn{1}{c}{$\nu$$^{\rm a}$} &
\multicolumn{1}{c}{$HPBW$} &
\multicolumn{1}{c}{$E_{\rm up}$$^a$} &
\multicolumn{1}{c}{$S\mu^2$$^a$} &
\multicolumn{1}{c}{rms} &
\multicolumn{1}{c}{$T_{\rm peak}$$^b$} &
\multicolumn{1}{c}{$V_{\rm peak}$$^b$} &
\multicolumn{1}{c}{$FWHM$$^b$} &
\multicolumn{1}{c}{$I_{\rm int}$$^b$} \\
\multicolumn{1}{c}{ } &
\multicolumn{1}{c}{(GHz)} &
\multicolumn{1}{c}{($\arcsec$)} &
\multicolumn{1}{c}{(K)} &
\multicolumn{1}{c}{(D$^2$)} & 
\multicolumn{1}{c}{(mK)} &
\multicolumn{1}{c}{(mK)} &
\multicolumn{1}{c}{(km s$^{-1}$)} &
\multicolumn{1}{c}{(km s$^{-1}$)} &
\multicolumn{1}{c}{(mK km s$^{-1}$)} \\ 

\hline
\multicolumn{10}{c}{DC$_{\rm 3}$N}\\
\hline


DC$_{\rm 3}$N 10--9, F=9--8 & 84.4298 &  29 & 22 & 124 &  \multirow{3}{*}{6}  &  \multirow{3}{*}{31 (1)}  &  \multirow{3}{*}{+8.72 (0.13)}     &  \multirow{3}{*}{1.0 (0.3)}    &  \multirow{3}{*}{34 (8)}   \\
DC$_{\rm 3}$N 10--9, F=10--9 & 84.4298 & 29 & 22 & 137\\
\vspace{0.25cm} \hspace{-0.2cm}
DC$_{\rm 3}$N 10--9, F=11--10 &84.4298 & 29 & 22 & 152 \\

DC$_{\rm 3}$N 11--10, F=10--9 & 92.8724 &  26 & 27 & 138 &  \multirow{3}{*}{2} &  \multirow{3}{*}{12 (1)} &  \multirow{3}{*}{+8.96 (0.10)} &  \multirow{3}{*}{1.6 (0.2)}  &  \multirow{3}{*}{21 (3)}   \\
DC$_{\rm 3}$N 11--10, F=11--10 & 92.8724  & 26 & 27 & 151 \\
\vspace{0.25cm} \hspace{-0.2cm}
DC$_{\rm 3}$N 11--10, F=12--11 & 92.8724  & 26 & 27 & 166 \\

DC$_{\rm 3}$N 12--11, F=11--10 & 101.3148 &  24 & 31 & 152 &  \multirow{3}{*}{1} &  \multirow{3}{*}{12 (1)} &  \multirow{3}{*}{+8.82 (0.02)}&  \multirow{3}{*}{1.9 (0.1)}  &  \multirow{3}{*}{24 (1)}   \\
DC$_{\rm 3}$N 12--11, F=12--11 & 101.3148  & 24 & 31 & 165\\
\vspace{0.25cm} \hspace{-0.2cm}
DC$_{\rm 3}$N 12--11, F=13--12 & 101.3148  & 24 & 31 & 180 \\

DC$_{\rm 3}$N 13--12, F=12--11 & 109.7571 &  22 & 37 & 166 &  \multirow{3}{*}{1} &   \multirow{3}{*}{12 (1)} &  \multirow{3}{*}{+8.62 (0.05)} &  \multirow{3}{*}{1.1 (0.1)}  &  \multirow{3}{*}{14 (1)}   \\
DC$_{\rm 3}$N 13--12, F=13--12 & 109.7571  & 22 & 37 & 179 \\
\vspace{0.25cm} \hspace{-0.2cm}
DC$_{\rm 3}$N 13--12, F=14--13 & 109.7571  & 22 & 37 & 194 \\

DC$_{\rm 3}$N 16--15, F=15--14 & 135.0832 &  18 & 55 & 208 & \multirow{3}{*}{1} &   \multirow{3}{*}{39 (1)}  &  \multirow{3}{*}{+8.40 (0.01)} &  \multirow{3}{*}{0.5 (0.1)}  &  \multirow{3}{*}{19 (1)}   \\
DC$_{\rm 3}$N 16--15, F=16--15 & 135.0832 & 18 & 55 & 221 \\
\vspace{0.25cm} \hspace{-0.2cm}
DC$_{\rm 3}$N 16--15, F=17--16 & 135.0832 & 18 & 55 & 235 \\

DC$_{\rm 3}$N 17--16, F=16--15 & 143.5249 &  17 & 62 & 222 & \multirow{3}{*}{3} &  \multirow{3}{*}{18 (3)} &  \multirow{3}{*}{+7.76 (0.06)} &  \multirow{3}{*}{0.7 (0.1)}  &  \multirow{3}{*}{14 (2)}   \\
DC$_{\rm 3}$N 17--16, F=17--16 & 143.5249 & 17 & 62 & 235 \\ 
\vspace{0.25cm} \hspace{-0.2cm}
DC$_{\rm 3}$N 17--16, F=18--17 & 143.5249 & 17 & 62 & 249 \\

DC$_{\rm 3}$N 18--17, F=17--16 & 151.9664 &  16 & 69 & 235 & \multirow{3}{*}{12} &  \multirow{3}{*}{20 (4)} &  \multirow{3}{*}{+8.94 (0.20)} &  \multirow{3}{*}{0.4 (0.7)}  &  \multirow{3}{*}{8 (6)}   \\
DC$_{\rm 3}$N 18--17, F=18--17 & 151.9664 & 16 & 69 & 249 \\
DC$_{\rm 3}$N 18--17, F=19--18 & 151.9664 & 16 & 69 & 263 \\
\hline
\multicolumn{10}{c}{HC$_{\rm 3}$N}\\
\hline

HC$_{\rm 3}$N 9--8 & 81.8815 & 30 & 20 & 125 & 9 & 642 (41) & +9.00 (0.01) & 1.4 (0.02) & 947 (13)\\
HC$_{\rm 3}$N 10--9 & 90.9790 & 27 & 24 & 139 & 3 & 613 (35) & +8.60 (0.01) & 1.5 (0.008) & 950 (4)\\
HC$_{\rm 3}$N 10--11 & 100.0764 &25  & 29 & 153 &  3 & 587 (30) & +8.43 (0.01) & 1.4 (0.008) & 880 (5)\\
HC$_{\rm 3}$N 12--11 & 109.1736 & 23 & 34 & 166 & 4 & 526 (17) & +8.46 (0.01) & 1.3 (0.01) & 747 (5)\\

HC$_{\rm 3}$N 15--14 & 136.4644 &18 & 52 & 208 & 9 & 562 (12) &  +8.16 (0.01) & 1.2 (0.02) & 745 (11)\\
HC$_{\rm 3}$N 16--15 & 145.5609 & 17 & 59 & 222 & 14 & 540 (18) & +8.34 (0.01) & 1.1 (0.03) & 654 (14)\\
HC$_{\rm 3}$N 17--16 & 154.6573 &16  & 67 & 236 & 9 & 426 (18) & +8.16 (0.01) & 1.2 (0.02) & 521 (9)\\
HC$_{\rm 3}$N 18--17 & 163.7534 & 15 & 75 & 250 & 8 & 380 (22) & +8.37 (0.01) & 1.3 (0.03) & 524 (9)\\

HC$_{\rm 3}$N 23--22 & 209.2302 &12  & 120 & 319 & 11 & 175 (10) & +8.47 (0.03) & 1.9 (0.08) & 353 (13)\\ 
HC$_{\rm 3}$N 24--23 & 218.3248 & 11 & 131 & 333 & 12 & 166 (12) & +8.64 (0.03) & 2.1 (0.1) & 375 (14)\\
HC$_{\rm 3}$N 25--24 & 227.4189 & 11 & 142 & 347 & 10 & 125 (7) & +8.50 (0.04) & 2.2 (0.1) & 292 (12)\\
HC$_{\rm 3}$N 26--25 & 236.5128 &10 & 153 & 361 & 10 & 87 (9) & +8.41 (0.05) & 2.1 (0.1) & 196 (11)\\
HC$_{\rm 3}$N 27--26 &  245.6063 & 10 & 165 & 374 & 11 & 66 (8) & +8.60 (0.09) & 2.8 (0.3) &  199 (15)\\


\hline
\noalign{\vskip 2mm}
\end{tabular}}

$^a$ Frequencies and spectroscopic parameters of DC$_{\rm 3}$N and HC$_{\rm 3}$N have been extracted from the Jet Propulsion Laboratory data base \citep{Pickett1998}. $^b$The errors in brackets are the Gaussian fit uncertainties. 
\label{Table:HC3N}
\end{table*}

\subsubsection{4.2.2 Thioformaldehyde}
\begin{figure}
\begin{center}
\includegraphics[angle=0,width=10cm]{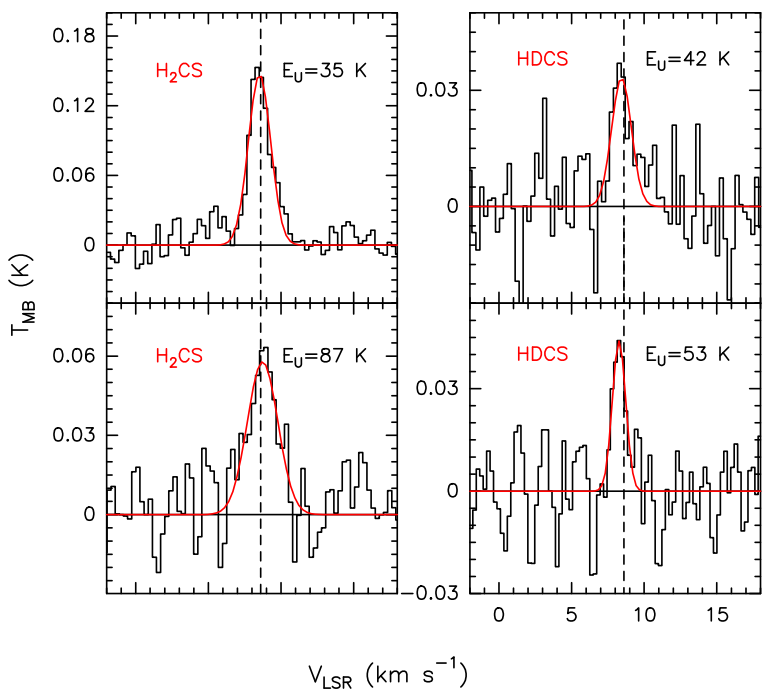} 
  \caption{Examples of the observed spectra of H$_{\rm 2}$CS and HDCS, in T$_{\rm MB}$ scale (i.e. not corrected for the beam dilution): species and excitation energies are reported in the upper left and right corner of each panel, respectively. The vertical dashed line stands for the ambient LSR velocity (+ 8.6 Km s$^{-1}$: \citep{Chen2009}).}
  \label{Fig:spectra-H2CS}
  \end{center}
\end{figure}

The ASAI spectra detected 11 lines of thioformaldehyde and 5 lines of its single deuterated isotopologue HDCS. The line excitation energies $E_{\rm up}$ are in the range 35--165 K and 31--64 K for H$_{\rm 2}$CS and HDCS, respectively.
Examples of the detected line spectra are shown in Figure \ref{Fig:spectra-H2CS}; the detected transitions and the observational parameters are listed in Table 2. The line profiles are close to a Gaussian shape and the peak velocities are close to the systemic source velocity, with values between +7.6 and +8.7 km s$^{-1}$ for H$_{\rm 2}$CS and between +7.8 and +8.4 km s$^{-1}$ for HDCS.
The line FWHM range is 1.8--3.8 km s$^{-1}$ for the main isotopologue and 0.7--2.3 km s$^{-1}$ for the single deuterated isotopologue.
All the transitions of H$_{\rm 2}$CS and HDCS are detected in the ASAI 1mm band, except for one transition of HDCS detected at 2mm.

\begin{table*}
\resizebox{\textwidth}{!}{
\caption{List of transitions and line properties (in $T_{\rm MB}$ scale) of the HDCS and H$_{\rm 2}$CS emission detected towards SVS13-A.}
\begin{tabular}{lccccccccc}
 \hline
\multicolumn{1}{c}{Transition} &
\multicolumn{1}{c}{$\nu$$^{\rm a}$} &
\multicolumn{1}{c}{$HPBW$} &
\multicolumn{1}{c}{$E_{\rm up}$$^a$} &
\multicolumn{1}{c}{$S\mu^2$$^a$} &
\multicolumn{1}{c}{rms} &
\multicolumn{1}{c}{$T_{\rm peak}$$^b$} &
\multicolumn{1}{c}{$V_{\rm peak}$$^b$} &
\multicolumn{1}{c}{$FWHM$$^b$} &
\multicolumn{1}{c}{$I_{\rm int}$$^b$} \\
\multicolumn{1}{c}{ } &
\multicolumn{1}{c}{(GHz)} &
\multicolumn{1}{c}{($\arcsec$)} &
\multicolumn{1}{c}{(K)} &
\multicolumn{1}{c}{(D$^2$)} & 
\multicolumn{1}{c}{(mK)} &
\multicolumn{1}{c}{(mK)} &
\multicolumn{1}{c}{(km s$^{-1}$)} &
\multicolumn{1}{c}{(km s$^{-1}$)} &
\multicolumn{1}{c}{(mK km s$^{-1}$)} \\ 

\hline
\multicolumn{10}{c}{HDCS}\\
\hline

HDCS 5$_{\rm 1,5}$--4$_{\rm 1,4}$ & 151.9275 & 16 & 31 & 13 &  7 & 31 (7) & +8.36 (0.11) & 1.1 (0.3) & 37 (7)\\
HDCS 7$_{\rm 1,7}$--6$_{\rm 1,6}$ & 212.6483 & 12 & 50 & 19 & 9 &  39 (4) & +7.81 (0.08) & 0.7 (0.2) & 29 (6)\\
HDCS 7$_{\rm 0,7}$--6$_{\rm 0,6}$ & 216.6624 & 12 & 42 & 19 & 10 &  33 (5) & +8.44 (0.18) & 1.6 (0.4) & 57 (12)\\
HDCS 8$_{\rm 0,8}$--7$_{\rm 0,7}$ & 247.4885 & 10 & 53 & 22 &  9 & 22 (3) & +8.27 (0.09) & 1.1 (0.2) & 51 (9)\\
HDCS 8$_{\rm 1,7}$--7$_{\rm 1,6}$ & 252.7350 & 10 & 64 & 21 & 3 & 27 (2) & +8.32 (0.05) & 2.3 (0.2) & 67 (3)\\
\hline
\multicolumn{10}{c}{H$_{\rm 2}$CS }\\
\hline

o-H$_{\rm 2}$CS 6$_{\rm 1,6}$--5$_{\rm 1,5}$ & 202.9241     & 12 & 47 & 47 &   24 & 222 (27) & +8.45 (0.06) & 2.0 (0.2) & 468 (33) \\
p-H$_{\rm 2}$CS 6$_{\rm 0,6}$--5$_{\rm 0,5}$ & 205.9879    & 12 & 35 & 16 &  27 & 147 (12) &  +8.53 (0.10) & 1.8 (0.3) & 283 (31) \\
p-H$_{\rm 2}$CS 6$_{\rm 2,4}$--5$_{\rm 2,3}$ & 206.1586     & 12 & 87 &  15 & 11 & 58 (10) & +8.74 (0.11) & 2.5 (0.3) & 153 (13)\\
o-H$_{\rm 2}$CS 6$_{\rm 1,5}$--5$_{\rm 1,4}$ & 209.2006      & 12 &  48 & 48 & 13 & 181 (12) & +8.37 (0.03) & 1.8 (0.1) & 351 (15)\\
o-H$_{\rm 2}$CS 7$_{\rm 1,7}$--6$_{\rm 1,6}$ & 236.7270       & 10 & 59 &  56 & 13 & 175 (13) & +7.82 (0.04) & 2.4 (0.1) & 448 (16)\\
p-H$_{\rm 2}$CS 7$_{\rm 0,7}$--6$_{\rm 0,6}$ & 240.2669     & 10 &  46 &19 & 12 & 110 (9) & +8.56 (0.06) & 2.8 (0.2) & 333 (16)\\
p-H$_{\rm 2}$CS 7$_{\rm 2,6}$--6$_{\rm 2,5}$ & 240.3821      & 10 & 99 & 17 & 30 & 74 (16) &  +8.05 (0.27) & 3.3 (0.6) & 262 (41)\\
o-H$_{\rm 2}$CS 7$_{\rm 3,5}$--6$_{\rm 3,4}$ & 240.3930     & 10 &  \multirow{2}{*}{165} &\multirow{2}{*}{47}  &\multirow{2}{*}{27}  & \multirow{2}{*}{90 (13)} & \multirow{2}{*}{+7.65(0.21)} & \multirow{2}{*}{3.8 (0.5)} & \multirow{2}{*}{362 (40)}\\
o-H$_{\rm 2}$CS 7$_{\rm 3,4}$--6$_{\rm 3,3}$ & 240.3938 & 10 \\
p-H$_{\rm 2}$CS 7$_{\rm 2,5}$--6$_{\rm 2,4}$ & 240.5491 & 10 &99 & 17 &  9 & 45 (11) & +8.16 (0.13) & 3.2 (0.3) & 153 (13)\\
o-H$_{\rm 2}$CS 7$_{\rm 1,6}$--6$_{\rm 1,5}$ & 244.0485 & 10 & 60 &56 &  10 & 178 (12) & +8.49 (0.03) & 2.9 (0.1) & 556 (13)\\
o-H$_{\rm 2}$CS 8$_{\rm 1,8}$--7$_{\rm 1,7}$ & 270.5219 & 9 & 72 & 64 & 15 & 124 (11) & +8.46 (0.07) & 2.9 (0.2) & 384 (18)\\
\hline
\noalign{\vskip 2mm}
\end{tabular}}

$^a$Frequencies and spectroscopic parameters of HDCS and H$_{\rm 2}$CS have been extracted from the Cologne Database for Molecular Spectroscopy \citep{Muller2005}.$^b$The errors in brackets are the gaussian fit uncertainties. 
\label{Table:H2CS}
\end{table*}

\subsection{4.3 Non-LTE analysis of the line emission}

We analysed the HC$_{\rm 3}$N and H$_{\rm 2}$CS observed lines via the non-LTE large velocity gradient (LVG) approach using the model described by Ceccarelli et al. \citep{Ceccarelli2003}.
For HC$_{\rm 3}$N, we used the collisional coefficients with H$_{\rm 2}$ from Faure et al. \citep{Faure2016}, provided by the BASECOL database \citep{Dubernet2013}. For H$_{\rm 2}$CS, we used the collisional coefficients with H$_{\rm 2}$ computed for the H$_2$CO--H$_2$ system by Wiesenfeld \& Faure  \citep{Wiesenfeld2013}, scaled for the different mass, and provided by the LAMDA database \citep{Schoier2005}.
In the calculations, we assumed a Boltzmann distribution for the H$_{\rm 2}$ ortho-to-para ratio.

We ran two large grids of models varying the gas temperature, T$_{\rm kin}$, from 20 to 200 K; the H$_{\rm 2}$ density, n$_{\rm H_{\rm 2}}$, from $1\times10^5$ to $2\times10^8$ cm$^{-3}$; the HC$_{\rm 3}$N column density, N(HC$_3$N), from $1\times10^{12}$ to $4\times10^{16}$ cm$^{-3}$;
the H$_{\rm 2}$CS column density, N(H$_2$CS), from $1\times10^{13}$ to $4\times10^{16}$ cm$^{-3}$. 
We then found the solution with the lowest $\chi^2$, in the temperature-density-column density-size space. 

\subsubsection{4.3.1 Cyanoacetylene}
\begin{figure}
\begin{center}
\includegraphics[angle=0,width=16cm]{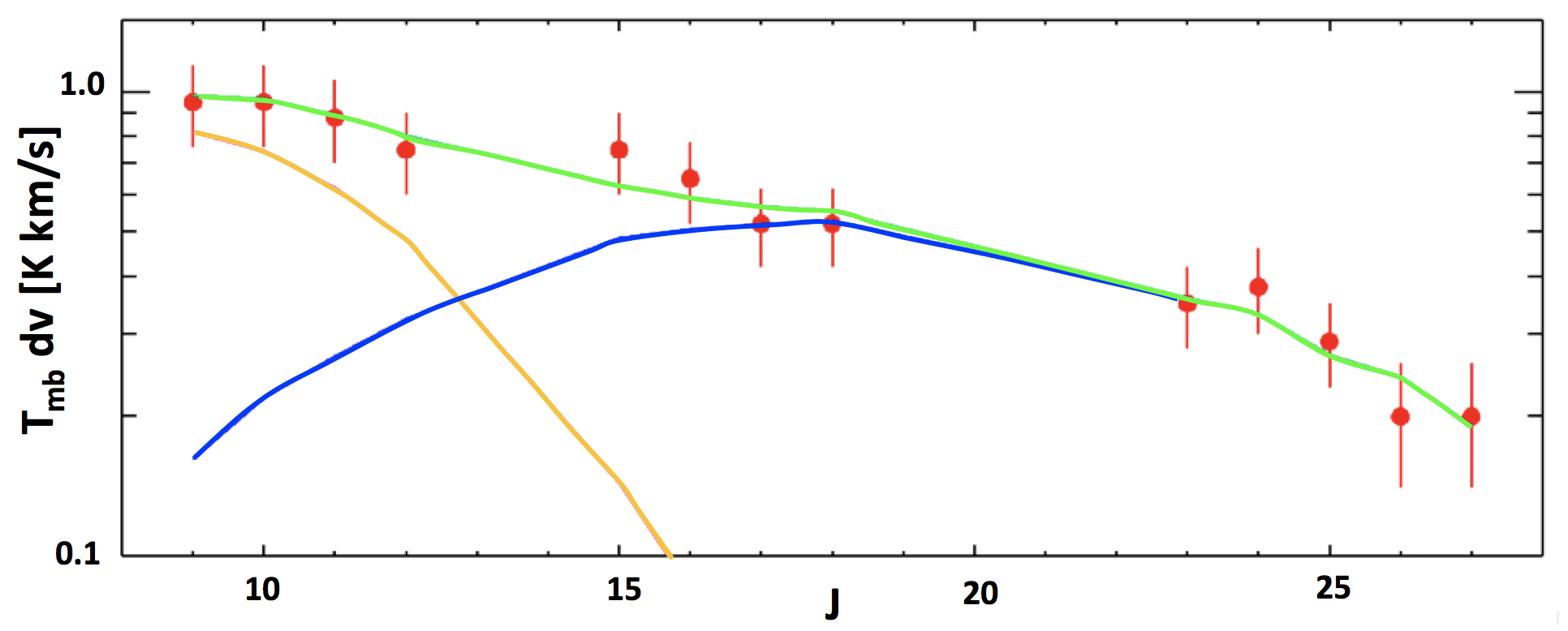} 
  \caption{Observed (red points) and theoretical (curves) line intensity of the H$_{\rm 3}$CN lines as a function of the quantum number J. The best fit (green curve) of the lines is obtained considering two different components: one extended region (45$\arcsec$) with N(H$_{\rm 3}$CN)=2$\times$10$^{12}$ cm$^{-2}$, T$_{\rm kin}$=20K--30 K and a density of 1$\times$10$^5$ cm$^{-3}$ (orange curve) and a second component with a size of 9$\arcsec$, N(H$_{\rm 3}$CN)=1$\times$ 10$^{13}$ cm$^{-2}$, T$_{\rm kin}$=40 K and a density of 8$\times$10$^6$ cm$^{-3}$ (blue curve).}
  \label{Fig:hc3n-LVG}
  \end{center}
\end{figure}

The Spectral Line Energy Distribution (SLED) of the detected lines is shown in Fig. \ref{Fig:hc3n-LVG}. It is obvious looking at the SLED that the emitting gas is composed of (at least) two components. So, we searched for a fit with two components. We obtained a good fit (see Fig. \ref{Fig:hc3n-LVG}) with the parameters reported in Table \ref{tab:HC3N-LVG} and that indicates the presence of a cold ($\sim$20 K) extended component, likely associated with the extended envelope and/or molecular cloud, and a second lukewarm ($\sim$40 K), more compact and denser one, whose origin will be discussed later.

\begin{table*}
\caption{Results from the non-LTE LVG analysis. Columns reports the derived gas temperatures, $T_{\rm kin}$, density $n_{\rm H_{\rm 2}}$, column densities, $N_{\rm tot}$, and size, $\theta$ for the two components traced by the HC$_3$N line emission.}
\begin{tabular}{cccc}
\hline
\hline
\multicolumn{1}{c}{$T_{\rm kin}$} &
\multicolumn{1}{c}{$n_{\rm H_{\rm 2}}$} &
\multicolumn{1}{c}{$N_{\rm tot}$} &
\multicolumn{1}{c}{$\theta$}\\
\multicolumn{1}{c}{(K) } &
\multicolumn{1}{c}{(cm$^{-3}$) } &
\multicolumn{1}{c}{(cm$^{-2}$)} &
\multicolumn{1}{c}{($\arcsec$)} \\
\hline
20 K --30 K & 1 $\times$ 10$^{5}$ & 2 $\times$ 10$^{12}$ & 45\\
40 K & 8 $\times$ 10$^{6}$ & 1 $\times$ 10$^{13}$ & 9\\
\hline
\label{tab:HC3N-LVG}
\end{tabular}
\end{table*}

\subsubsection{4.3.2 Thioformaldehyde}
\begin{figure}
\begin{center}
\includegraphics[angle=0,width=8.cm]{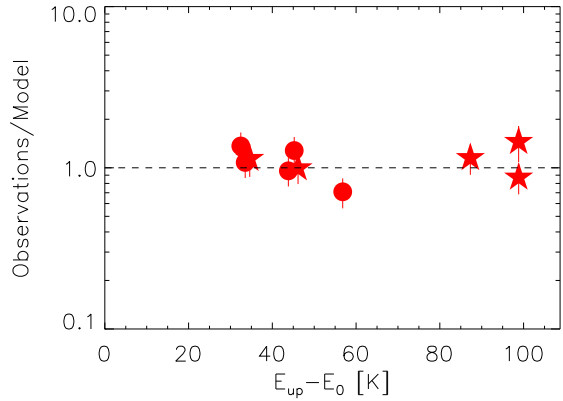} 
\caption{Ratio between the observed and the best-fit theoretical line intensities of H$_2$CS as a function of the line upper level energy (see text). Stars and circles indicate para- and ortho- transitions, respectively.}
  \label{Fig:h2cs-LVG}
  \end{center}
\end{figure}

In the case of H$_2$CS, one component fits very well with all the measured line intensities, as shown in Fig. \ref{Fig:h2cs-LVG}. The H$_{\rm 2}$CS column density and emitting size are well constrained: N(H$_{\rm 2}$CS)=6$\times$ 10$^{15}$ cm$^{-2}$ and $\theta$=0$\farcs$6 (equivalent to $\sim$120 au). On the contrary, the density and temperature are degenerate and not well constrained: both a 40--60 K and n$_{\rm H_{\rm 2}}\geq 8\times 10^4$ cm$^{-3}$ or a 60--100 K and n$_{\rm H_{\rm 2}}\leq 3\times 10^5$ cm$^{-3}$ component give the same $\chi^2$. The origin of this component will be discussed in Section 5.
Finally, we notice that, in order to have a good fit, the H$_{\rm 2}$CS ortho-to-para ratio has to be equal to 3, namely the statistical value.

\subsubsection{4.3.3 Deuterated cyanoacetylene and thioformaldehyde}
We derived the deuteration of cyanoacetylene and thioformaldehyde by dividing the intensities of lines with the same J number for DC$_3$N/HC$_3$N and similar upper level energy, E$_{\rm up}$, for HDCS/H$_2$CS \citep{Kahane2013}. This method has a twofold advantage with respect to the commonly used one that consists in dividing column densities derived by rotational diagrams: it allows us to determine the deuteration with a (much) lower error and as a function of the energy of the transition, which can then provide precious information on the deuteration in the different components of the emitting gas.
Please note that we corrected the intensities of the main isotopologues to account for the optical depths estimated from the LVG analysis, described in two previous sections. Similarly, we took into account the H$_{\rm 2}$CS ortho-to-para ratio derived in the previous section.

\begin{figure}
\begin{center}
\includegraphics[angle=0,width=10cm]{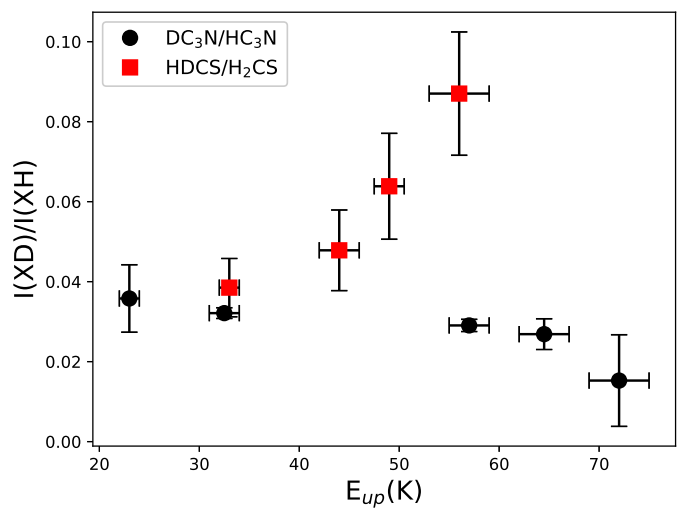} 
  \caption{Deuteration ratio of HC$_3$N (black points) and H$_{\rm 2}$CS (red squares) as a function of the line upper level energy E$_{\rm up}$. Please note that the E$_{\rm up}$ of the line is linked to its excitation conditions so that larger E$_{\rm up}$ likely trace warmer and/or denser gas.}
  \label{Fig:H2CS-deut}
  \end{center}
\end{figure}

Figure \ref{Fig:H2CS-deut} shows the deuteration of the two molecules, cyanoacetylene and thioformaldehyde, as a function of the upper level energy of the transition E$_{\rm up}$, namely their excitation conditions: roughly, the larger E$_{\rm up}$ the warmer and/or denser the emitting gas.
The molecular deuteration varies from $\sim$3.5\% to $\sim$1.5\% in HC$_3$N with increasing E$_{\rm up}$, while it goes from $\sim$4\% to $\sim$9\% in H$_2$CS. Therefore, it seems that while the molecular deuteration decreases in warmer/denser HC$_3$N emitting gas, it increases in the H$_2$CS emitting gas. We will discuss the possible meaning of this different behavior in Section 5.

\section{5. Discussion}
\subsubsection{5.1 Deuterated cyanoacetylene and thioformaldehyde across SVS13-A}

In the previous section, we have shown a puzzling, at first sight, behavior of the deuteration of cyanoacetylene and thioformaldehyde: while the former decreases with increasing upper level energy of the lines E$_{\rm up}$, namely the excitation conditions of the lines, the latter decreases with increasing E$_{\rm up}$. This is, at first sight, rather puzzling. However, we will show here that this is completely explicable when one considers the physical and chemical structure of SVS13-A. Therefore, we will start from that.

\subsubsection{5.1.1 Physical and chemical structure of SVS13-A}
The analysis of our new observations reported in Section 4.3 helps to understand the physical and chemical structure of SVS13-A, its hot corino and surrounding envelope.
The hot corino, namely the region with temperatures greater than about 100 K, has a radius of about 45 au and is enriched with HDO, CH$_3$OH and other iCOMs. The density of the hot corino is not well constrained, only a lower limit of $\sim 10^8$ cm$^{-3}$ is obtained.
On the other hand, the HC$_3$N line emission identifies an extended region, with a radius of about $\sim$7000 au, with a temperature of about 20 K and density of about $10^5$ cm$^{-3}$.
In between these two regions, there is the lukewarm envelope with intermediate temperatures, $\sim$ 40 K, and densities, $\sim 10^6-10^7$ cm$^{-3}$, traced by H$_2$CO and HC$_3$N.
Very likely, the innermost part of the lukewarm envelope is the region where H$_2$CS emits the lines that we observe.

The intriguing behavior of the HC$_3$N and H$_2$CS deuteration as a function of increasing line excitation conditions, shown in Fig. \ref{Fig:H2CS-deut}, can now be interpreted in the light of the structure described in Section 5.1 and provides information on the history of SVS13-A.

\subsubsection{5.1.2 DC$_3$N/HC$_3$N} 
Cyanoacetylene is located in the cold and lukewarm envelope and it is, extremely likely, formed by gas-phase reactions occurring in the present (mostly from the reaction of C$_2$ with CN) \citep{Wakelam2015,Jaber2017}. Therefore, it is totally natural that lines with increasing E$_{\rm up}$, which probe gas with increasing temperatures, show a decreasing molecular deuteration as they reflect the deuteration caused by the present gas-phase H$_2$D$^+$/H$_3^+$ abundance ratio, which decreases with increasing temperature \citep{Ceccarelli2014}. We emphasize the relatively large value of the measured DC$_3$N/HC$_3$N ratio (0.01--0.04), which points to a substantial depletion of gaseous CO in the envelope of SVS13-A \citep{Roberts2000,Bacmann2003}. Indeed, gas-phase CO destroys the H$_2$D$^+$ ion that is responsible for the formation in the gas-phase of most deuterated species.

\subsubsection{5.1.3 HDCS/H$_2$CS} 
Thioformaldehyde is present in the innermost regions of the lukewarm envelope, at about 40--60 K. It is very likely produced on the grain-surfaces by the hydrogenation of frozen CS, even though no laboratory experiments nor theoretical calculations exist in the literature to substantiate this hypothesis. The CS ($\sim 0.77$ ea$_a$) dipole moment is larger than that of CO ($\sim 0.39$ ea$_0$), although the electronegativity of atomic S is closer to that of C than O\footnote{Interestingly, though, the CS dipole moment enhancement comes from the contribution of both the charges on each atom and the induced dipoles on either C and S atoms which add up instead of partially cancelling, as it is the case for CO \citep{Harrison2006}.}. This probably implies that CS gets adsorbed on the icy surfaces of dust grains much easier than CO, and, hence, that H$_2$CS is formed relatively fast. H$_2$CS would then be injected into the gas-phase when the dust temperature reaches the iced-H$_2$CS sublimation temperature. Unfortunately, no laboratory measurements of the binding energy of H$_2$CS on icy surfaces are available. The only theoretical estimate, based on the interaction of one molecule of H$_2$CS with two water molecules rather than the whole ice structure, gives a value of $\sim$3000 K \citep{wakelam2017}. This would correspond to a sublimation of frozen H$_2$CS when the dust temperature reaches about 40---60 K, which is surprisingly consistent with the region where we detect H$_2$CS in SVS13-A. 

We can now try to interpret why HDCS/H$_2$CS increases in warmer/denser regions, as shown by Fig. \ref{Fig:H2CS-deut}.
In general, the molecular deuteration of hydrogenated species on the grain-surfaces increases with decreasing gaseous CO abundance at the time of ice formation, namely with increasing density \citep{Taquet2012b}. We showed above that the envelope of SVS13-A is centrally peaked, namely the density increases going inward. There is no reason to think that this was not the case at the time of the iced H$_2$CS formation. Therefore, going inwards corresponds to ices that are more and more deuterated, which is what we see in the HDCS/H$_2$CS abundance ratio.

\subsubsection{5.2 Comparison with other sources}
Figure \ref{Fig:Deut} summarises the deuteration found in different molecules and different sources. In this section, we will focus on the cyanoacetylene and thioformaldehyde deuteration, the other species having been discussed in Section 3.2.

\subsubsection{5.2.1 DC$_3$N/HC$_3$N} 
The value of cyanoacetylene deuteration measured in SVS13-A (0.01--0.04) is similar to that found from previous measurements towards cold cores (0.02--0.1) \citep{Langer1980, Howe1994}. It is also similar to that found towards the first hydrostatic core Chamaeleon-MMS1 ($\sim$0.04) \citep{Cordiner2012} and in the Class 0 protostar L1527 ($\sim$0.03) \citep{Sakai2009b}.

More recently, cyanoacetylene deuteration has been measured towards the Class 0 protostar IRAS16293-2422 \citep{Jaber2017}, where it was found that DC$_3$N/HC$_3$N varies from 0.5 in the cold envelope to 0.05 in the hot corino region. As in the case of SVS13-A, the deuteration in the cold envelope is likely a present-day gas-phase product. The larger measured DC$_3$N/HC$_3$N value indicates a more substantial depletion of gaseous CO in IRAS16293-2422 than in SVS13-A (see above), which is in agreement with SVS13-A being more evolved (namely with a globally warmer envelope) than IRAS16293-2422. 
The absence of HC$_3$N in the hot corino of SVS13-A, contrarily to IRAS16293-2422, maybe due to a slower formation of the prestellar core from which SVS13-A originates, following the arguments presented by Jaber et al. \citep{Jaber2017}.

\subsubsection{5.2.2 HDCS/H$_2$CS} 
The thioformaldehyde deuteration measured in SVS13-A (0.04--0.10) is not very different, within the errors, to that measured in IRAS 16293-2422 (HDCS/H$_{\rm 2}$CS=0.1$\pm$0.014 \citep{Droz2018}), the only Class 0 protostar where HDCS/H$_{\rm 2}$CS has been also measured.
The deuteration of H$_{\rm 2}$CS has been previously measured in the dark cloud Barnard 1 ($\sim$ 0.3 \citep{Marcelino2005}), and in the prestellar core L1544 \citep{Vastel2018}. The measured value is 0.20$\pm$0.14, again similar, within the error bars, to that measured towards SVS13-A. 

\subsubsection{5.2.3 Conclusive remarks}
Looking at Fig. \ref{Fig:Deut}, there does not seem to exist a dramatic decrease of cyanoacetylene and thioformaldehyde going from the prestellar core to the Class I protostar phases. 
However, from the discussion above it is clear that the information is too scarce to be able to assess whether the lack of a trend can be linked to a lack of evolution in the deuteration of cyanoacetylene and thioformaldehyde with time. In addition, it is also clear that the specific conditions of each object where the deuteration is measured enter into the equation and could even be the dominant effect, as is the case when comparing, for example, the DC$_3$N/HC$_3$N in the Class 0 protostar IRAS16293-2422 and the Class I protostar SVS13-A.

\section{6. Conclusions and perspectives}
In this contribution, we argued that Class I protostars are the bridge between the youngest Class 0 protostar phase and the more evolved Class II/III, where it is believed planets and comets form in their protoplanetary disks. Analogously, we argued that Class I protostars are crucial objects to study the chemical evolution throughout the solar-type star forming process and to understand what is passed from one stage to the successive one.

We then presented a review of the chemical properties of Class I, specifically focused on the iCOMs and deuterated molecules and showed that most of the information about Class I protostars comes from studies towards SVS13-A. The first obvious conclusion, therefore, is that the information is extremely scarce and that many more observations are necessary to carry out a meaningful comparison among different evolutionary stages.

Having said that, our preliminary conclusions are:
\begin{itemize}
    \item The complete census of iCOMs emission performed in SVS13-A suggests that the chemical richness of Class I protostars is comparable to that of Class 0 sources. 
    \item The comparison of the deuteration of methanol seems rather to indicate a decrease in Class I sources. However, we cautioned on the large uncertainties linked to the limited observations available.
    \item The new observations presented here allow us  to reconstruct the physical and chemical structure of SVS13-A, which is composed by a hot corino and an envelope extending up to $\sim7000$ au in radius. The envelope has an increasing density and temperature going inward. From a chemical point of view, the outermost region is probed by HC$_3$N emission, then by H$_2$CO, and finally by H$_2$CS.
    \item The DC$_3$N/HC$_3$N abundance ratio decreases going inward whereas that of HDCS/H$_2$CS increases. We explained this behavior with the different chemical formation of cyanoacetylene and thioformaldehyde: the former is a present-day gas-phase product whereas the latter is a past grain-surface product.
\end{itemize}

Obviously, more observations of Class I protostars are necessary to draw a reliable picture of the chemical evolution during the solar-type formation process. This means not only a much larger number of sources, but also better accuracy and spatial resolution, so to be able to disentangle the various components making a Class I protostar. 
Some observational programs are already in progress, like the IRAM-NOEMA Large Program SOLIS ({\it Seeds Of Life In Space}) \citep{Ceccarelli2017} and the ALMA Large Program FAUST (Fifty AU STudy of the chemistry in the disk/envelope system of Solar-like protostars; http://stars.riken.jp/faust/fausthome.html, PI: Yamamoto). Both projects aim at collecting much more observational data towards Class I protostars, in addition to other sources in different evolutionary phases.

In addition, an important question remains about the opacity of the iCOMs line in the millimeter, which affects their abundance measure.
An illustrative example of this problem is represented by the HH212 Class 0 protostar, which was imaged with ALMA with a resolution of 10 au \citep{Lee2017c, Codella2018, Lee2019}. While iCOMs are detected in the upper atmosphere of its circumstellar disk, no emission is found towards the denser disk midplane \citep{Lee2017a}. The question is: is the lack of iCOMs emission towards the midplane due to an intrinsic small abundance or is it caused by a high continuum optical depth? To answer this question observations in the cm-domain are needed. Therefore, the future SKA telescope will be instrumental in this respect (see e.g. Cradle of Life\citep{Codella2015SKA} at https://astronomers.skatelescope.org/science-working-groups/galaxy-cradle-life/).

We would also like to emphasize the importance of the synergy between astronomers and chemists for the interpretation of observations. The case of H$_2$CS described in Section 5 provides an illustrative example on the necessity to have laboratory experiments and theoretical quantum chemistry calculations in order to be able to draw firm conclusions from the astronomical observations. We can mention the large-scale project ACO (Astro-Chemical Origins: https://aco-itn.oapd.inaf.it/home, PI: Ceccarelli, grant No 811312), funded by the European Commission, which gathers together instrumentalists, astronomers, chemists and computer scientists.

\begin{acknowledgement}

This work is based on observations carried out with the IRAM-NOEMA Interferometer and IRAM-30m telescope. IRAM is supported by INSU/CNRS (France), MPG (Germany) and IGN (Spain).
This project has received funding from (i) the European Research Council (ERC) under the European Union's Horizon 2020 research and innovation programme, for the Project "The Dawn of Organic Chemistry" (DOC), grant agreement No 741002 and (ii) the European MARIE SK\L{}ODOWSKA-CURIE ACTIONS under the European Union's Horizon 2020 research and innovation programme, for the Project "Astro-Chemistry Origins" (ACO), Grant No 811312.
This work was supported by the PRIN-INAF 2016 "The Cradle of Life - GENESIS-SKA (General Conditions in Early Planetary Systems for the rise of life with SKA)".

\end{acknowledgement}


\setkeys{acs}{articletitle=true}
\bibliography{MyBib}



\end{document}